\documentclass[10pt,twocolumn,reqno, amsmath,amssymb, aps,prb,superscriptaddress]{revtex4-1}

\usepackage{float}
\usepackage{perpage}

\usepackage{color}
\usepackage{rotating}
\usepackage[breaklinks=true,colorlinks,citecolor=blue,linkcolor=blue,urlcolor=blue]{hyperref}
\usepackage{appendix}
\usepackage{graphicx}
\usepackage{natbib}
\usepackage{dcolumn}
\usepackage{bm}
\usepackage{hyperref}
\usepackage[mathlines]{lineno}
\usepackage[caption=false]{subfig}
\usepackage{epstopdf}

\begin{document}

\title{Optomechanical response of a non-linear mechanical resonator}
\author{Olga Shevchuk}
\affiliation{Kavli Institute of Nanoscience, Delft University of Technology, Lorentzweg 1, 2628 CJ Delft, The Netherlands}

\author{Vibhor Singh}
\affiliation{Kavli Institute of Nanoscience, Delft University of Technology, Lorentzweg 1, 2628 CJ Delft, The Netherlands}
\author{Gary A. Steele}
\affiliation{Kavli Institute of Nanoscience, Delft University of Technology, Lorentzweg 1, 2628 CJ Delft, The Netherlands}
\author{Ya. M. Blanter}
\affiliation{Kavli Institute of Nanoscience, Delft University of Technology, Lorentzweg 1, 2628 CJ Delft, The Netherlands}

\begin{abstract}
We investigate theoretically in detail the non-linear effects in the response of an optical/microwave cavity coupled to a Duffing mechanical resonator. The cavity is driven by a laser at a red or blue mechanical subband, and a probe laser measures the reflection close to the cavity resonance. Under these conditions, we find that the cavity exhibits optomechanically induced reflection (OMIR) or absorption (OMIA) and investigate the optomechanical response in the limit of non-linear driving of the mechanics. Similar to linear mechanical drive, an overcoupled cavity the red-sideband drive may lead to both OMIA and OMIR depending on the strength of the drive, whereas the blue-sideband drive only leads to OMIR. The dynamics of the phase of the mechanical resonator leads to the difference between the shapes of the response of the cavity and the amplitude response of the driven Duffing oscillator, for example, at weak red-sideband drive the OMIA dip has no inflection point. We also verify that mechanical non-linearities beyond Duffing model have little effect on the size of the OMIA dip though they affect the width of the dip.
\end{abstract}
\maketitle

\section{Introduction}
In optomechanics, light and mechanical motion are coupled together by the radiation pressure of photons trapped in an optical cavity. Optomechanics has the potential to provide access to the quantum limit of mechanical motion, and in recent years, has been a field that has been undergoing rapid development~\cite{Aspelmeyer}. It already resulted in a number of ground-breaking experiments, including, for example, observation of radiation pressure shot noise \cite{Purdy} or optomechanical squeezing of light \cite{Safavi,Purdy1}. Following the first observation of quantum nature of nanomechanical resonator \cite{Cleland}, experiments in microwave optomechanical  \cite{Teufel} and cavity optomecanical \cite{Painter,Verhagen} architectures demonstrated that mechanical resonators can be brought to the quantum regime. The next step will be to achieve reliable quantum manipulation and to use mechanical resonators as quantum memory elements and quantum transducers.

An important tool in optomechanics is optomechanically induced transparency (OMIT), an effect analogous to electromagnetically induced transparency in quantum optics \cite{Scully,Agarwal}. In an OMIT experiment, the cavity is strongly driven by a drive (pump) laser at the red-sideband --- the driving frequency is red-shifted from the cavity resonance by the frequency of the mechanical resonator --- and the transmission through the cavity is measured by a probe laser close to the cavity resonance. Due to the interference, the transmission exhibits a narrow peak exactly at the cavity resonance. The width of the peak is determined by the mechanical relaxation rate, and the OMIT peak is typically much more narrow than the cavity resonance. The observation of the OMIT peak serves as the signature of optomechanical coupling and can be used to qualitatively determine the coupling in the experiment. OMIT was first observed in optomehanical experiments in Ref. \onlinecite{Weiss}. It has been heavily used for characterization of optomechanical systems, in particular, for extraction of the value of optomechanical couplingmechanical resonators and optomechanical coupling, in both optical \cite{Weiss,Painter,Hocke} and microwave \cite{Teufel1,Vibhor}
realizations. In single-port cavities, like the one used in Ref. \onlinecite{Vibhor}, one can only measure reflection. A peak/dip in the reflection coefficient is referred to as optomechanically induced reflection/absorption (OMIR/OMIA) \cite{Hocke}. 

In most optomechanical experiments, both cavity and mechanical resonator were linear. However, non-linear effects are conceptually important. In the classical regime, a driven non-linear oscillator exhibits bistable behavior, which may strongly affect the properties of the system. Moreover, in the quantum regime, non-linear effects are essential for creation of non-classical states of a non-linear resonator. There are three sources of non-linear behavior in cavity and microwave optomechanics. First, the radiation pressure interaction is inherently non-linear, but typically a cavity is strongly driven by a laser, and in this case the interaction can be linearized \cite{Aspelmeyer}. There were theoretical proposals to use the radiation pressure interaction for quantum manipulation of mechanical resonators \cite{Borkje}, which require strong coupling regime of operation. Second, a cavity can be made non-linear. Whereas it is difficult to achieve for optical cavities, microwave cavities can be made non-linear by adding Josephson junctions. Self-sustained oscillations caused by non-linear Lorentz force backaction \cite{Etaki} and cavity-induced Duffing behavior of a mechanical resonator \cite{Buks} were demonstrated in Josephson-based devices the dc regime. Third, mechanical resonators are inherently non-linear. This is best manifest in low-dimensional resonators such as carbon nanotubes and graphene flakes \cite{Bachtold}. In this Article, we concentrate on non-linear effects in mechanical system.

Non-linear mechanical effects in optomechanical architecture were experimentally studied in Ref. \onlinecite{Zhou} (see also Ref. \onlinecite{Hocke2014}). These articles investigated OMIT in an undercoupled optical cavity and discovered that the OMIT feature is still in place, and the shape of the OMIT peak repeats the shape of the frequency response of the driven non-linear (Duffing) oscillator and exhibits bistability resulting in a hysteretic behavior. 

Recently, we performed experimental studies of non-linear effects in OMIA \cite{exp} in the same overcoupled microwave cavity as Ref. \onlinecite{Vibhor}. The results we found have a qualitatively different lineshape from that of Ref. \onlinecite{Zhou} --- the shape of the OMIA dip in the reflection coefficient of the cavity was distorted by non-linear effects, however, the shape is different from the response of a driven Duffing oscillator, in particular, it does not have an inflection point.

In this Article, we provide a theoretical analysis of non-linear effects on OMIR/OMIA feature and explain the difference between Refs. \onlinecite{Zhou} and \onlinecite{exp}. We demonstrate that whereas the OMIR/OMIA feature persists in all situations and exhibits the bistability at strong enough probe powers, its shape is different depending on how the cavity is coupled to an external circuit. We show that the non-Duffing shape of the OMIA response is due to the effect of the phase of the mechanical resonator imprinted on the microwave field. We also analyze the OMIR/OMIA for a blue-detuned drive. 

The paper is organized as follows. In Section \ref{sec:system}, we develop a general theory of non-linear effects in OMIR/OMIA based on input-output relations --- a standard technique in quantum optics, and derive the OMIA for a red-sideband driven overcoupled cavity. In Section \ref{sec:Detuning} we generalize the results to the case when the drive is detuned from the red sideband, and in Sec. \ref{sec:Diagram} we demonstrate what happens for an undercoupled cavity, connecting our results with Ref. \onlinecite{Zhou}, and for a blue-sideband drive. In Section \ref{sec:beta}, we consider non-linear effects beyond the Duffing appoximation: The quadratic term in the force and the non-linear dissipation. Section \ref{sec:conclusion} presents the conclusions.

\section{Model of a driven non-linear cavity}
\label{sec:system}

\begin{figure}
\centering
\includegraphics[width=0.4\textwidth]{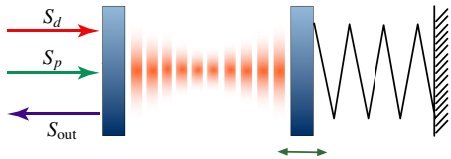}
\caption{Schematic of an optomechanical cavity consisting of a static (left) and a movable (right) mirrors. The movable mirror acts as a mechanical resonator. It is coupled to the cavity through the radiation pressure.}
\label{fig:Optsys}
\end{figure}

We consider a single-port cavity, which is coupled via the radiation pressure to a non-linear resonator as shown in Fig. \ref{fig:Optsys}. The Hamiltonian of the system is 
\begin{eqnarray}
\hat{H}=\hbar \omega_c \hat{a}^\dagger \hat{a}+\frac{\hat{p}^2}{2 m_{eff}}+\frac{m_{eff} \Omega_m^2 \hat{x}^2}{2}+\alpha\frac{\hat{x}^4}{4}\nonumber\\
+\hbar G \hat{x} \hat{a}^\dagger \hat{a} + i\hbar \sqrt{\eta \kappa}\left[ s_{in}(t)\hat{a}^\dagger-s_{in}^*(t)\hat{a} \right] \ .
\label{Duffing_model}
\end{eqnarray}
Here $\hat{a}^\dagger(\hat{a})$ is the creation (annihilation) operator for the cavity mode with the cavity frequency $\omega_c$. The resonator with the effective mass $m_{eff}$ and the resonance frequency $\Omega_m$ is described by the operators of mechanical displacement $\hat{x}$ and momentum $\hat{p}$ operators. The mechanical non-linearity is described by the Duffing term with the strength $\alpha$. The radiation pressure term contains the optomechanical coupling constant  $G=d\omega_c/dx$ between the mechanical and cavity modes, which corresponds to the shift of the cavity frequency due to the motion of the resonator. The last term describes the coupling of the input signal to the cavity. The cavity coupling parameter is given by $\eta=\kappa_e/(\kappa_0+\kappa_e)=\kappa_e/\kappa$, where $\kappa_0$, $\kappa_e$ and $\kappa$ denote the intrinsic, the external, and the total dissipation rates, respectively.  The input signal $s_{in}(t)=S_d e^{-i\omega_d t}+S_p e^{-i\omega_p t} $ consists of the strong drive field with the normalized field amplitude $S_d$ and drive frequency $\omega_d$ and weak probe field with the normalized amplitude $S_p$ and probe frequency $\omega_p$. The amplitudes  $S_d$ and $S_p$ are square roots of the corresponding drive/probe power divided by $\hbar \omega_c$ and have dimensions of $s^{-1/2}$. $S_p^2$ and $S_d^2$ have the meaning of the incident photon flux measured in photons per second. In the remainder of the article we will refer to the amplitudes $S_d$ and $S_p$ simply as drive and probe fields.

Writing the Heisenberg equations of motion in the rotating-wave frame of the drive frequency\cite{Aspelmeyer,Weiss} and adding dissipation of the cavity and mechanical modes, we derive the quantum Langevin equations for our system as follows,
\begin{eqnarray}
&&\frac{d}{dt} \hat{a}(t)=\left(i\Delta -\frac{\kappa}{2}\right)\hat{a}(t) -iG\hat{x}(t) \hat{a}(t)+\sqrt{\eta\kappa}s_{in}(t)e^{i\omega_d t},\quad
\\
&&m_{eff}\left(\frac{d^2}{dt^2} \hat{x}(t)+\Gamma_m\frac{d}{dt} \hat{x}(t)+\Omega_m^2 \hat{x}(t)\right)
\nonumber 
\\
&& \qquad {} \qquad \qquad \qquad \qquad=-\hbar G \hat{a}^\dagger(t) \hat{a}(t) -\alpha\hat{x}^3(t), 
\end{eqnarray} 
where $\Delta=\omega_c-\omega_d$ is the detuning between the cavity field and drive field and $\Gamma_m$ is the damping rate of the resonator.

In the steady-state, disregarding the probe field, the time derivatives vanish, and the static solutions for the intra-cavity field and mechanical displacement obey the following algebraic equations,
\begin{eqnarray}
&&\overline{a}=\frac{\sqrt{\eta\kappa}}{-i\overline{\Delta}+\kappa/2}S_d,
\\
&&m_{eff}\Omega_m^2\overline{x}+\alpha\overline{x}^3+\hbar G \overline{a}^2=0,
\end{eqnarray}
where $\overline{\Delta}=\Delta-G\overline{x}$ is an effective cavity detuning including the frequency shift due the static mechanical displacement.

Since the probe field is  much weaker than the drive field, following the standard methods of quantum optics, we can rewrite each Heisenberg operator as the sum of its steady-state mean value and a small fluctuation, which has zero mean value
\begin{equation}
\hat{a}(t)=\overline{a}+\delta\hat{a}(t)\hspace{5pt}\text{and}\hspace{5pt}\delta\hat{x}(t)=\overline{x}+\delta\hat{x}(t). 
\end{equation}
In this case, the steady-state values are governed by the drive power and the small fluctuations by the probe power. Then, keeping only the linear terms of the fluctuation operators in the radiation pressure term (disregarding $\delta \hat{a}^\dagger(t) \delta \hat{a}(t) $ and $\delta \hat{a}(t)\delta\hat{x}(t) $), we obtain the linearized quantum Langevin equations  
\begin{eqnarray}
&&\frac{d}{dt} \delta\hat{a}(t)=(i\overline{\Delta}-\frac{\kappa}{2}) \delta\hat{a}(t)-iG\overline{a}\delta\hat{x}(t)+\sqrt{\eta\kappa}S_p e^{-i\Omega t},\quad
\label{a_eqn}
\\
&& m_{eff}\left(\frac{d^2}{dt^2} \delta\hat{x}(t)+\Gamma_m\frac{d}{dt} \delta\hat{x}(t)+\Omega_m^2\delta\hat{x}(t)\right)=-\hbar G\overline{a}(\delta\hat{a}(t)
\nonumber
\\
&&+\delta\hat{a}^\dagger (t))-\alpha(\delta\hat{x}^3(t)+3\overline{x}^2\delta\hat{x}(t)+3\overline{x}\delta\hat{x}^2(t)),\label{x_eqn}\quad
\end{eqnarray}
 with $\Omega=\omega_p-\omega_d$ being the detuning of the probe field from the drive. In order to solve this system of equations we introduce the following Ansatz:  $\delta\hat{a}(t)=Ae^{-i\Omega t}+A^+ e^{+i\Omega t}$. We are interested here in the resolved sideband regime ($\kappa\ll\Omega_m$) and close to the blue ($\overline{\Delta}=\Omega_m$) or red ($\overline{\Delta}=-\Omega_m$) sideband of the drive field, meaning the lower sideband $A^+$ is far off-resonance and can be neglected. Upon substituting this Ansatz into Eq. \eqref{a_eqn} we derive the amplitude $A$ of the cavity field 
\begin{equation}
A=\frac{-iG\overline{a} e^{i\Omega t}\delta\hat{x}}{-i(\overline{\Delta}+\Omega)+\frac{\kappa}{2}}+\frac{\sqrt{\eta\kappa} S_p}{-i(\overline{\Delta}+\Omega)+\frac{\kappa}{2}}.
\end{equation}

Next, we substitute this into  Eq. \eqref{x_eqn}. This is a non-linear equation which can be solved in the same way as one solves the forced Duffing oscillator \cite{Strogatz}. We rewrite the equation in terms of the dimensionless time $\tau=\Omega_m t$ and frequency $\omega=\Omega/\Omega_m$
\begin{eqnarray}
&&\frac{d^2}{d\tau^2} \delta\hat{x}(\tau)+\omega^2\delta\hat{x}(\tau)=-(\omega^2-1)\delta\hat{x}(\tau)-\frac{\Gamma_m}{\Omega_m}\frac{d}{d\tau} \delta\hat{x}(\tau)
\nonumber
\\
&&\qquad-\frac{\hbar G\overline{a} }{m_{eff}\Omega_m^2 }(A e^{i\omega\tau}+A^* e^{-i\omega\tau})
\nonumber
\\
&&\qquad-\frac{\alpha}{m_{eff}\Omega_m^2}(\delta\hat{x}^3(\tau)+3\overline{x}^2\delta\hat{x}(\tau)+3\overline{x}\delta\hat{x}^2(\tau)).\quad
\label{x_eqn_tau}
\end{eqnarray}
The terms proportional to $\overline{x}^3$ and $\overline{x}\delta\hat{x}^2(\tau)$ are dropped since they create a very small static force.  To facilitate analysis of this non-linear equation, the small asymptotic parameter $\varepsilon\ll 1$ is introduced. Then, assuming weak non-linearity, weak damping, weak forcing with frequency close to the natural frequency of the mechanical resonator, one can perform transformations: $\alpha/m_{eff}\Omega_m^2\rightarrow\varepsilon\alpha/m_{eff}\Omega_m^2$, $\Gamma_m/\Omega_m\rightarrow\varepsilon \Gamma_m/\Omega_m$, $\hbar G\overline{a} A/m_{eff}\Omega_m^2\rightarrow\varepsilon \hbar G\overline{a} A/m_{eff}\Omega_m^2$ and $(\omega^2-1)\rightarrow\varepsilon(\omega^2-1)$. This indicates that all parameters on the r.h.s. of eq. \eqref{x_eqn_tau} are small, and we can construct the solution as a  power series using the method of multiple scales
\begin{equation}
\delta\hat{x}(\tau)=x_0(t_0,t_1)+\varepsilon x_1(t_0,t_1)+\mathcal{O}(\varepsilon^2),
\end{equation}
with the fast timescale $t_0=\tau$ and the slow  timescale $t_1=\varepsilon\tau$.  Applying two timing leads to the transformation of the first and second time derivative with respect to time $\tau$ 
\begin{eqnarray}
&&\frac{d}{d\tau}=\frac{\partial}{\partial t_0}+\varepsilon\frac{\partial }{\partial t_1}=D_0+\varepsilon D_1,
\\
&&\frac{d^2}{d\tau^2}= D_0^2+2\varepsilon D_1D_0+\mathcal{O}(\varepsilon^2).
\end{eqnarray}
Using relations for the derivatives in the equation of motion for mechanical resonator and collecting orders $\mathcal{O}(1)$ and $\mathcal{O}(\varepsilon)$ yield the pair of differential equations
 \begin{eqnarray}
&& D_0^2x_0+\omega^2x_0=0,
\\ 
&&D_0^2x_1+\omega^2x_1=\left(\omega^2-1+\frac{i\hbar G^2\overline{a}^2/m_{eff} \Omega_m^2}{-i(\overline{\Delta}+\omega\Omega_m)+\frac{\kappa}{2}}\right)x_0
\nonumber
\\
&&-2 D_1D_0 x_0-\frac{\Gamma_m}{\Omega_m} D_0 x_0 -\frac{\hbar G \overline{a} \sqrt{\eta\kappa} S_p e^{-i\omega t_0}/m_{eff}\Omega_m^2}{-i(\overline{\Delta}+\omega\Omega_m)+\frac{\kappa}{2})}
\nonumber
\\
&&\hspace{110 pt}-\frac{\alpha}{m_{eff}\Omega_m^2}(x_0^3+3\overline{x}^2 x_0).
\label{eq_x_1}
 \end{eqnarray}
The first equation is a simple harmonic oscillator with the general solution $x_0(t_0,t_1)=X(t_1)e^{-i\omega t_0}+X^*(t_1)e^{i\omega t_0}$. Substituting $x_0$ to the r.h.s. of eq. \eqref{eq_x_1} and collecting terms proportional to the $e^{-i\omega t_0}$ we obtain the force that drives the l.h.s. at its resonance frequency. In order for the perturbation correction $x_1$ to not diverge, the sum of all collected terms must be zero. It gives rise to the condition for the slowly varying amplitude $X(t_1)$,
\begin{eqnarray}
&&2i \omega\dot{X}+\left(\omega^2-1+\frac{i\hbar G^2\overline{a}^2/m_{eff}\Omega_m^2}{-i(\overline{\Delta}+\omega\Omega_m)+\frac{\kappa}{2}}\right)X
\nonumber
\\
&&+i\omega \frac{\Gamma_m}{\Omega_m} X-\frac{\hbar G \overline{a} \sqrt{\eta\kappa} S_p/m_{eff}\Omega_m^2}{-i(\overline{\Delta}+\omega\Omega_m)+\frac{\kappa}{2}}
\nonumber
\\
&&-3\frac{\alpha}{m_{eff}\Omega_m^2}(X X^*+\overline{x}^2 )X=0.
\label{X_full}
\end{eqnarray}
This is our main result, which in the remainder of the Article we will simplify looking specifically at the blue or the red sideband of the cavity field.

\subsection{Red sideband}
For the red sideband, the drive field is fixed close the lower motional sideband $\overline{\Delta}=-\Omega_m+\delta$, where $\delta$ is a small shift (detuning) from the mechanical frequency. The probe field is slightly detuned from the cavity resonance frequency, $\Omega=\Omega_m+\Delta'-\delta$, where $\Delta'$ is a small detuning. We also define $g^2=\hbar G^2\overline{a}^2/2 m_{eff}\Omega_m$ and $F=\hbar G \overline{a} \sqrt{\eta\kappa} S_p/m_{eff}\Omega_m$. Then Eq. \eqref{X_full} is approximated as follows
\begin{eqnarray}
\label{X_eqn}
&&2i\dot{X}+2\left(\Delta' -\delta+\frac{i\Gamma_m}{2}+\frac{ig^2}{-i\Delta'+\frac{\kappa}{2}}\right)X
\nonumber
\\
&&-\frac{F}{-i\Delta'+\frac{\kappa}{2}}-3\frac{\alpha}{m_{eff}\Omega_m}(X X^*+\overline{x}^2 )X=0. \qquad
\end{eqnarray}
Introducing a polar form of the complex amplitude $X(t_1)=\frac{1}{2} b(t_1) e^{-i\phi(t_1)}$ and separating real and imaginary part ($2 i\dot{X}=i\dot{b}+b\dot{\phi}$) gives a system of two first-order differential equations for the real amplitude $b$ and phase $\phi$. These equations describe a slow time evolution of the amplitude and phase. To find the equilibrium points of this slow flow we look at the solution with constant amplitude($\dot{b}=0$) and phase($\dot{\phi}=0$)
\begin{eqnarray}
&&F\Delta'\sin(\phi)-\frac{\kappa}{2}F\cos(\phi)=g^2b\Delta'
\nonumber
\\
&&+\left(\frac{\kappa^2}{4}
+\Delta'^2\right)\left(\frac{3\alpha}{2m_{eff}\Omega_m}(b^2/4+\overline{x}^2 )-\Delta'+\delta\right) b,\qquad
\label{ampl_phase1}
\\
&&F\frac{\kappa}{2}\sin(\phi)+F\Delta'\cos(\phi)
\nonumber
\\
&&\hspace{70 pt}
=\left[\left(\frac{\kappa^2}{4}+\Delta'^2\right)\frac{\Gamma_m}{2}+g^2\frac{\kappa}{2}\right]b.
\label{ampl_phase2}
\end{eqnarray}

\begin{figure}[H]
\centering
\includegraphics[width=0.4\textwidth]{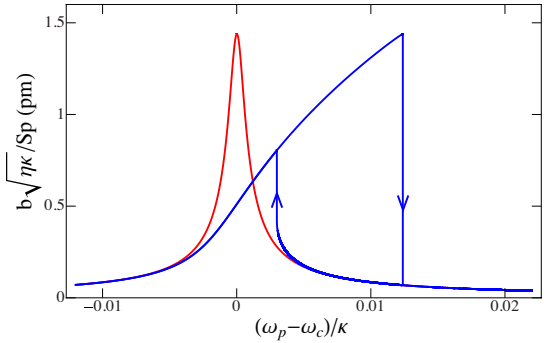}
\caption{(Color online) The amplitude of the mechanical resonator rescaled by the probe field $S_p= 10^4 s^{-1/2}$(red) and $S_p= 3\times 10^6 s^{-1/2}$(blue) as a function of the frequency $\Delta'/\kappa$ driven exactly at the red sideband. For the non-linear frequency response (blue) the hysteresis occurs. The blue arrow up shows the direction of the jump in the amplitude for decreasing frequency and the blue arrow down shows the direction of the jump for increasing frequency. Most parameters are taken from the experimental paper by Singh {\em et. al.}\cite{Vibhor} with the cavity coupling parameter $\eta=0.777$, except for the intra-cavity number of photons $\overline{a}^2=7.6\times 10^6$, mechanical damping $\Gamma_m/2\pi=  \hspace{3 pt} 200\hspace{3 pt}  Hz$, and non-linearity stregth $\alpha=2\times 10^{14}\hspace{5 pt}kg\hspace{3 pt}m^{-2} s^{-2}$. }
\label{fig:osc}
\end{figure}

The results for the amplitude of the mechanical resonator is shown on Fig. \ref{fig:osc} at zero detuning of the drive ($\delta =0$) and $\alpha > 0$ as a function of the probe frequency for two probe powers. We see that the behavior is exactly the same as  for the Duffing oscillator\cite{Strogatz}. At low driving (low probe power) the response of the amplitude is the same as for a linear oscillator and results in a Lorentzian peak. If the probe power increases, the resonant curve becomes asymmetric and develops instability, shown as hysteresis. 

The amplitude of the cavity field can now be found using real amplitude and phase of the mechanical oscillator,
\begin{equation} \label{ampl_gen}
A=\frac{-iG\overline{a} b e^{-i\phi}/2 }{-i(\overline{\Delta}+\Omega)+\frac{\kappa}{2}}+\frac{\sqrt{\eta\kappa} S_p}{-i(\overline{\Delta}+\Omega)+\frac{\kappa}{2}}.
\end{equation}

\subsection{Blue sideband}
For the blue sideband, we follow the same steps as for the red sideband by taking the frequencies $\Omega=\omega_p-\omega_d=-\Omega_m+\Delta'-\delta$ and $\overline{\Delta}=\omega_d-\omega_c-G\overline{x}=\Omega_m+\delta$. Consequently, the equations for the amplitude \eqref{ampl_phase1}  and phase \eqref{ampl_phase2}  can be adapted by changing sign in the term on the r.h.s. of Eq. \eqref{ampl_phase1} by $-\left(\frac{\kappa^2}{4}
+\Delta'^2\right)b\Delta'\rightarrow +\left(\frac{\kappa^2}{4}
+\Delta'^2\right) b\Delta'$ and  in Eq. \eqref{ampl_phase2} by $\left(\frac{\kappa^2}{4}
+\Delta'^2\right) \Gamma_m b/2\rightarrow-\left(\frac{\kappa^2}{4}
+\Delta'^2\right) \Gamma_m b/2$.

\subsection{Reflection coefficient}
In order to study OMIA or OMIR, we need to look at the cavity reflection of the probe field. It is defined by the ratio of the output and input field amplitudes at the probe frequency.  The output field of a single port cavity can be found using the input-output relationship
\begin{equation}
s_{out}(t)=s_{in}(t)+\sqrt{\eta \kappa} a(t).
\end{equation}
Hence, the reflection coefficient of the probe field is given by
\begin{eqnarray}
&&\vert S_{11}\vert =\left\vert 1-\frac{\sqrt{\eta \kappa} A}{S_p}\right\vert.
\end{eqnarray}
For a linear mechanical resonator, the  amplitude $A$ is linear with the probe field $S_p$, because the fluctuations of the mechanical resonator $\delta \hat{x}$ are linear with $S_p$. Hence, the reflection coefficient is independent of the probe field. For a non-linear resonator the relation is more complex.

At zero detuning $\delta=0$ and for a linear resonator, the shape of OMIT as the function of the detuning between drive and probe exactly repeats the shape of the response of the amplitude of the mechanical resonator. We see from Eqs. (\ref{ampl_phase1}), (\ref{ampl_phase2}) that this is not the case for a non-linear resonator. The main reason is that OMIT is not only determined by the dynamics of the amplitude $b$, but also by the dynamics of the phase shift $\phi$, which in turn depends on the frequency. This dependence is essentially an effect of back-action of the cavity on the resonator.

\begin{figure}[H]
\centering
\includegraphics[width=0.4\textwidth]{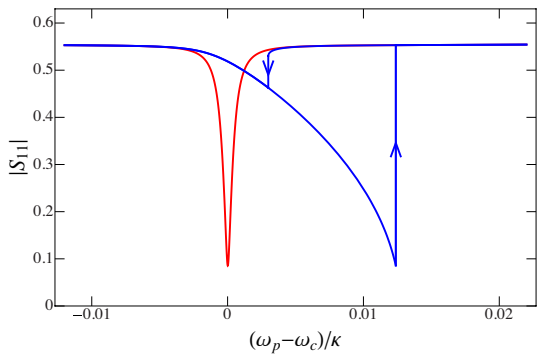}
\caption{(Color online) Optomechanically  induced absorption window found from the cavity reflection coefficient $|S_{11}|$, which corresponds to the mechanical amplitude shown in Fig. \ref{fig:osc}.}
\label{fig:OMIA}
\end{figure}

The non-trivial shape of reflection is illustrated in Fig. \ref{fig:OMIA} for the same parameters as Fig. \ref{fig:osc}. Note that in the remainder of the Article we will continue to use the same parameters to visualize the analysis unless mentioned otherwise. Only stable parts of the curve are shown; the bistability results in jumps between different branches of the reflection. It is seen that the reflection close to the cavity resonance is suppressed, indicating OMIA for a single-port cavity. The shape of the OMIA dip for stronger probe (Duffing oscillator) is bistable, however, it does not correspond to the shape of the resonator response (Fig. \ref{fig:osc}), for instance, it does not have an inflection point. 

\begin{figure}
\centering
{\includegraphics[width=0.4\textwidth]{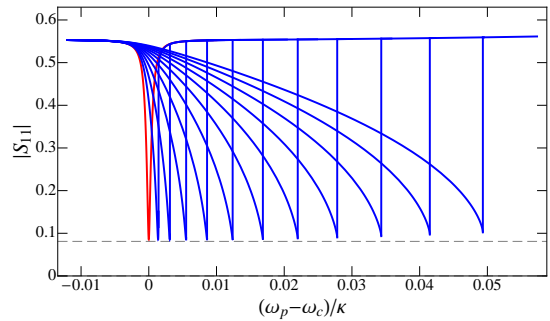}}\\
\caption{(Color online) OMIA response for different probe fields. The first red curve shows linear response at the probe field $S_p=10^4  s^{-1/2}$.  The dashed grey line represents the peak height of the linear response, in order to compare it with other probe fields. The blue curves start on the left from the probe field $S_p=1\times 10^6 s^{-1/2}$ and equally increase the probe field by  $S_p=0.5\times 10^6 s^{-1/2}$ step up to the last curve on the right, which corresponds to $S_p=6\times 10^6 s^{-1/2}$.}
\label{fig:dif_probe}
\end{figure}

Fig. \ref{fig:dif_probe} illustrates the dependence of the OMIA dip on the probe field $S_p$ for the same parameters as in Fig. \ref{fig:osc}. The red curve corresponds to the field $S_p=10^4  s^{-1/2}$, when the dynamics of the mechanical resonator is linear, and is the same as the red curve in Fig. \ref{fig:OMIA}. The blue curves, from left to right, correspond to increasing probe power, and the resonator shows non-linear behavior. We see that upon increasing the probe frequency, the width of the OMIA dip increases and the depth slightly decreases with the probe power. It does not correspond to the Duffing oscillator feature, where the peak height stays unaffected by the drive power.

In the rest of the Article, we investigate the effect of different parameters at the response of the cavity.

\section{OMIA detuning}
\label{sec:Detuning}

\begin{figure}
\centering
{\includegraphics[width=0.45\textwidth]{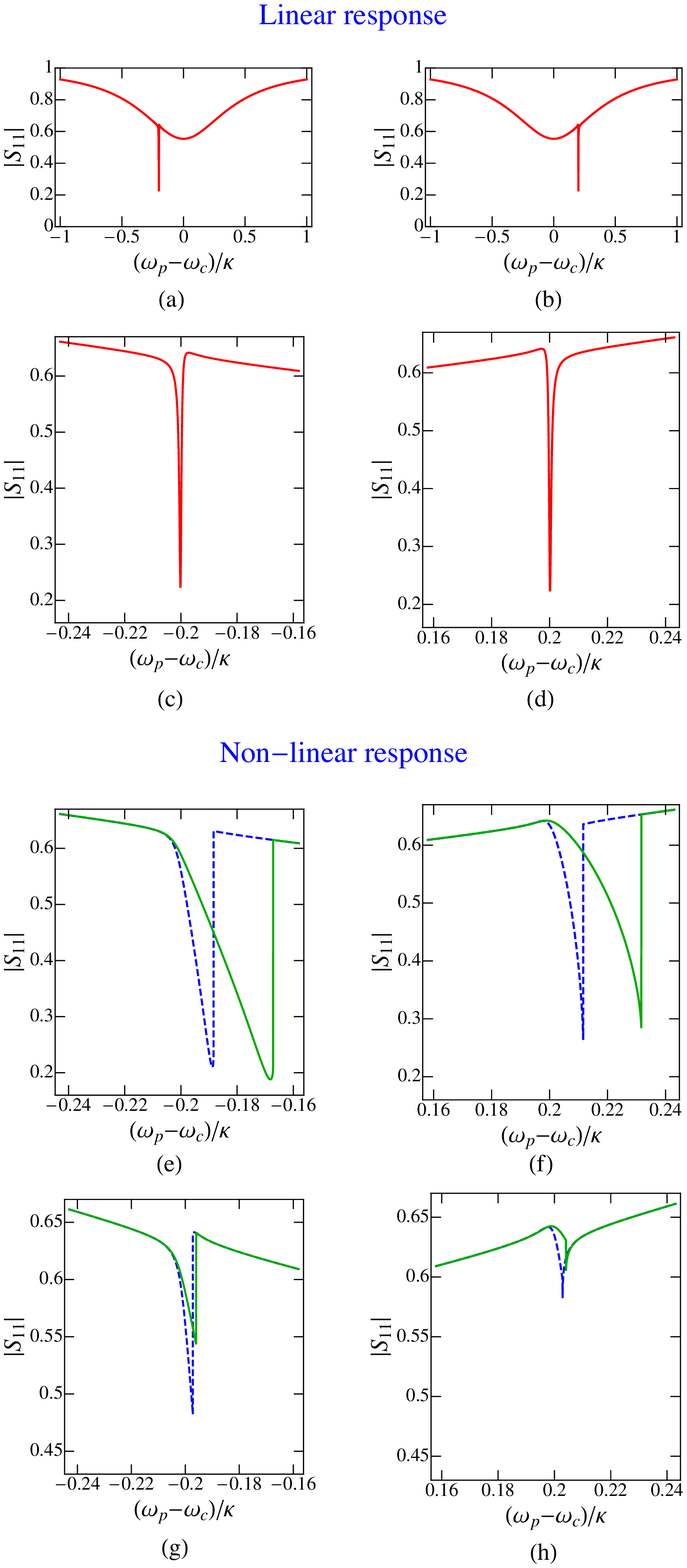}}\\
  \caption{ OMIA at non-zero detuning. All panels on the left show OMIA for the negative detuning $\delta=-0.2 \kappa$; all panels on the right are  for the same parameters but the positive detuning  $\delta=0.2 \kappa$. (a), (b) The full cavity reflection coefficient in the broad frequency range for the low probe field $S_p=10^4  s^{-1/2}$, when the cavity can be considered to be linear. (c), (d) Zoom of the linear OMIA peak. (e), (f) Non-linear OMIA window of the forward frequency sweep for the two probe fields:  $S_p=3\times 10^6  s^{-1/2}$ (dashed blue line) and $S_p=5\times 10^6  s^{-1/2}$ (solid green line). (g), (h) Non-linear OMIA of the backward frequency sweep for the same probe fields as in (e) and (f).}
  \label{fig:omia_det}
\end{figure}

In this Section, we look at the effect of the detuning of the drive $\delta$.  The results are shown in Fig. \ref{fig:omia_det} for the same parameters as Fig. \ref{fig:osc} and various values of $\delta$.

Fig. \ref{fig:omia_det} a and b give the general picture of the cavity response in a broad frequency window for positive (a) and negative (b) detunings at weak probe field, $S_p = 10^4  s^{-1/2}$. The detuning breaks the symmetry of the response with respect to the center of the cavity resonance even in the linear case, shifting the OMIA dip. We see indeed that the drive detuning shifts the narrow OMIA dip away from the minimum of the broad cavity resonance. This can be also seen from Eq. (\ref{ampl_phase1}): Indeed, $\delta$ enters there in the combination $\delta + 3 \alpha b^2/(8m_{eff} \Omega_m)$. We have seen already that increasing the probe power (increasing $b$) shifts the OMIA minimum to the right, therefore positive $\delta$ must shift it to the right as well, and the negative detuning must shift the peak to the left.

Fig. \ref{fig:omia_det} c and d show the structure of the dip for the same parameters as a and b, respectively. The shape is cleary asymmetric and is different from the Lorentzian shown in Fig. \ref{fig:OMIA}. The flank which is closer to the center of the cavity resonance (the right/left one for negative/positive detuning $\delta$) is sharper than the opposite flank. The OMIA dip at the detuning $\delta$ is a mirror reflection of the OMIA dip at the opposite detuning $-\delta$.

Fig. \ref{fig:omia_det} e, f and g, h show the cavity response at a stronger probe fields, $S_p=3\times 10^6  s^{-1/2}$ and $S_p=5\times 10^6  s^{-1/2}$, when the non-linear nature of the response is well pronounced. First, OMIA dips at $\delta$ and $-\delta$ are no more mirror images of each other. For the forward sweep in e and f, if the non-linear term is positive, $\alpha > 0$, the OMIA dip at negative detuning $-\delta$ is less pronounced than the dip for positive detuning $\delta$. On the contrary, for the backward sweep in g and h, the increase in the probe power results in the decrease of the depth for both, positive and negative, detunings.  For a linear system, we would expect a symmetry of the signal $S_{11} (\omega_p - \omega_c; \delta) = S_{11} (-\omega_p + \omega_c; -\delta$), which would make left and right panels of Fig. \ref{fig:omia_det} mirror reflections of each other.  In the presence of non-linearity, the curves are pulled in the same direction. Thus, the symmetry is broken, and therefore the OMIA dip is pronounced for the negative detuning. 

\section{Non-linear response map}
\label{sec:Diagram}
\begin{figure*}
\includegraphics[width= \linewidth]{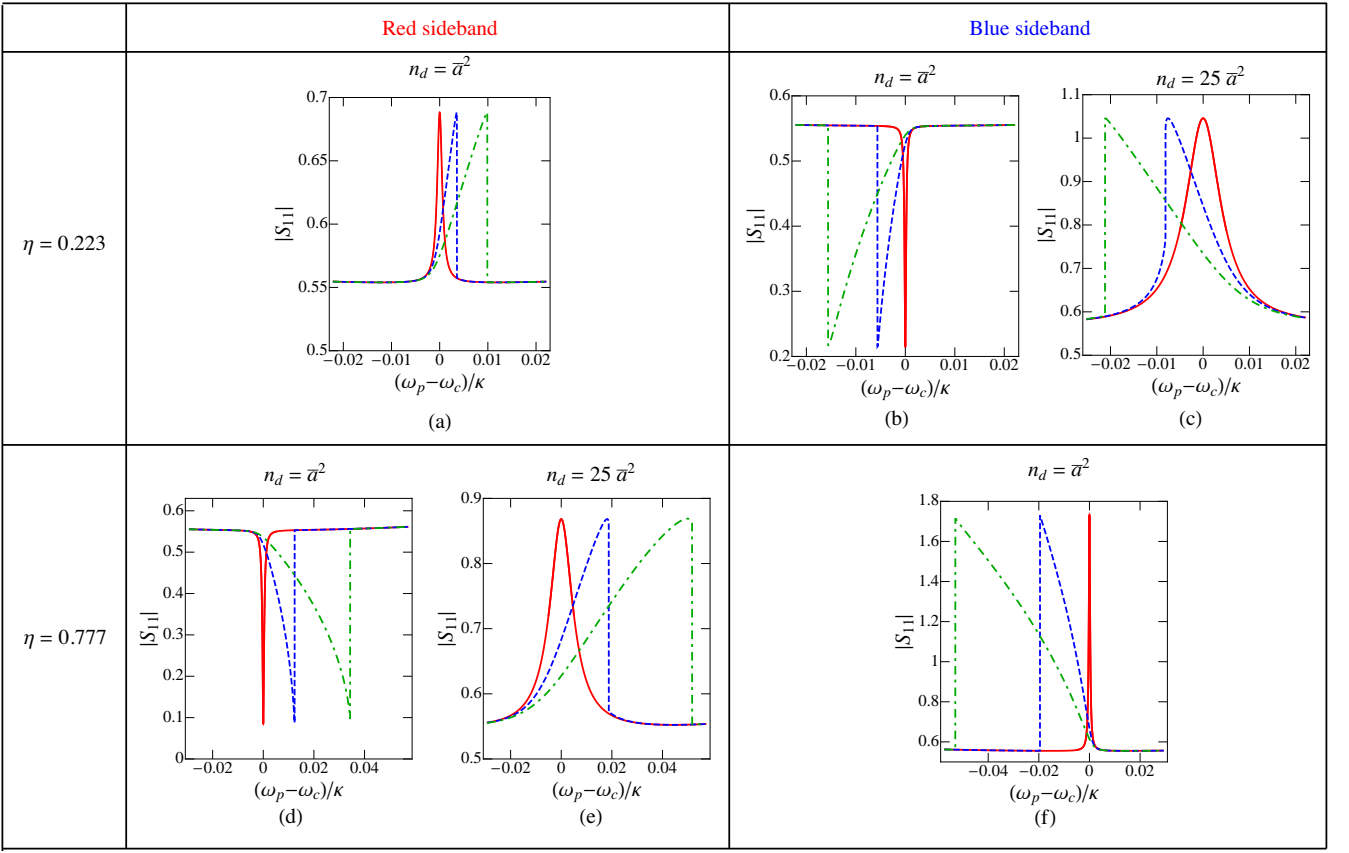}
\caption{Non-linear response map schematically shows the position and the shape of the OMIA/OMIR features for the undercoupled($\eta=0.223$), in the first row, and overcoupled($\eta=0.777$), in the second row, cavity. The two columns represent red and blue sideband.  When the intra-cavity number of photons $n_d=\overline{a}^2$ ( in the plots (a), (b), (d), and (f) ) the reflection coefficient is shown for the probe field $S_p=10^4 s^{-1/2}$ with the solid(red) line, $S_p=3\times 10^6  s^{-1/2}$ with the dashed(blue) line and $S_p=5\times 10^6  s^{-1/2}$ with the dash-dotted(green) line. For the larger intra-cavity number of photons $n_d=25 \overline{a}^2$ (in the plots (c) and (e) ) there is a switch between OMIA and OMIR , then the reflection coefficient is plotted for the following probe fields  $S_p=10^4  s^{-1/2}$ with the solid(red) line, $S_p=3\times 10^7 s^{-1/2}$ with the dashed(blue) line and $S_p=5\times 10^7  s^{-1/2}$ with the dash-dotted(green) line. Each plot corresponds to the different value of the Duffing non-linearity stregth $\alpha=2$ (a,d), $0.5$ (b,f), $0.08$ (c,e) $\times 10^{14} \hspace{5 pt}kg\hspace{3 pt}m^{-2} s^{-2}$. }
  \label{fig:omia_diag}
\end{figure*}

So far, we concentrated on red-sideband drive and overcoupled cavity, $\eta > 1/2$, and a moderate drive power. In this situation there is an OMIA dip which is again shown in the left curve of the left bottom cell of Fig. \ref{fig:omia_diag}. If the drive power increases, the cavity response $S_{11}$ instead of a minimum shows a maximum. This is an OMIR peak, as demonstrated in the right panel of the same cell of Fig. (\ref{fig:omia_diag}. Similarly for OMIA, at weak probe powers the OMIR peak is Lorentzian, and if we increase the probe power, the peak shifts to the right and broadens. Note that whereas the shape of the OMIA dip for weaker drive is different from the response of the mechanical resonator (no inflection point), the shape of the OMIR peak for stronger drive repeats the shape of the resonator response.

In the bottom right cell of Fig. \ref{fig:omia_diag} we show the results for the same overcoupled cavity driven at the blue sideband. We see that the OMIR peak develops at any drive power. For low probe powers the peak is Lorentzian, and for higher probe powers it shifts to the left and broadens. The shape of the peak is related to the shape of the response of the mechanical resonator, Fig. \ref{fig:osc}.


The qualitative difference between red- and blue-shifted drive for OMIR/OMIA has already been analyzed in Ref. \onlinecite{Vibhor} for the linear dynamics of the mechanical resonator. Indeed, the minimum value of the reflection coefficient is a non-monotonous function of the coupling parameter $\eta$: It decreases with $\eta$ for an undercoupled cavity, reaches zero for an optimally coupled cavity $\eta = 1/2$, and increases with $\eta$ for an overcoupled cavity. The red/blue-sideband drive modifies the cavity linewidth $\kappa_0$ such that it becomes $\kappa_0 \pm 4g^2/\Gamma_m$, where the upper/lower sign corresponds to the red/blue-detuned drive. Thus, for an overcoupled cavity increasing the drive intensity (proportional to the number of photons in the cavity) for the red-sideband drive takes the cavity towards the undercoupled limit, and the behavior changes dramatically when the effective coupling crosses the point $\eta = 1/2$, crossing over from OMIA to OMIR. In contrast, the blue-sideband drive only takes the cavity to even stronger overcoupled regime, and there are no qualitative changes. For an undercoupled cavity, the roles of red- and blue-sideband driving are swapped.

Now we turn to the non-linear dynamics. From Eq. (\ref{ampl_phase1}) we see that the signs of $b^2$ and $\Delta'$ are opposite for the red-sideband drive and the same for the blue-sideband drive. This means that since the OMIR peak shifts to the right from the cavity resonance for the red sideband, it shifts to the left for the blue sideband, in full accordance with Fig. \ref{fig:omia_diag}.

The top row of Fig. \ref{fig:omia_diag} shows OMIA/OMIT for an undercoupled cavity, $\eta < 1/2$. The red-sideband drive (top left corner) was previously studied in the experiments \cite{Zhou}. For any drive powers, there is an OMIR peak, and the shape of the peak correponds to the shape of the Duffing resonator response. The top right corner shows the cavity response for the blue-sideband drive. It is similar to what happens in an overcoupled cavity for the red-sideband drive: At low probe powers, one has an OMIA dip with the line shape different from the Duffing response of the resonator (no inflection point), whereas for stronger driving, an OMIR peak develops with the Duffing-like shape. The position of the peak is shifted to the left of the cavity resonance with increasing the probe power.

\section{Beyond Duffing}
\label{sec:beta}

To complete our results, we consider two more non-linear effects not included in the Duffing model (\ref{Duffing_model}) --- the quadratic term in the displacement and the non-linear damping. Both can be included in the treatment of Section \ref{sec:system}. 

\begin{figure}
\centering
\includegraphics[width=0.4\textwidth]{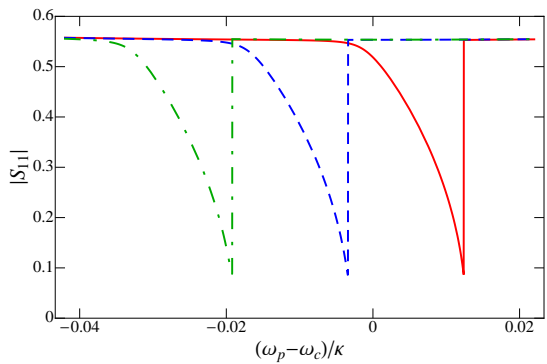}
\caption{ Evaluating the contribution of $\beta x^2$ to the energy of the mechanical resonator on the OMIA dip. The quadratic non-linearity $\beta=0$ is shown by solid(red) line, $\beta=3\times 10^{11} \hspace{5 pt}kg\hspace{3 pt}m^{-1} s^{-2}$ by dashed(blue) line, and $\beta=6\times 10^{11}  \hspace{5 pt}kg\hspace{3 pt}m^{-1} s^{-2}$ by dash-dotted(green) line for the fixed probe field $S_p=3\times 10^6  s^{-1/2}$. The $\beta x^2$ term induces a force-dependent shift of the equilibrium frequency, but otherwise does not change the non-linear dynamics or optomechanical response.}
\label{fig:beta}
\end{figure}

One phenomenon is the additional term $\beta x^2$ in the energy of the mechanical resonator. It produces the force proportional to the coordinate $x$ breaking the symmetry of the oscillator potential and shifting the equilibrium point. Here, we restrict ourselves to the case $\beta > 0$. Fig. \ref{fig:beta} shows the results for the overcouples cavity driven exactly at the red-sideband, $\delta = 0$. The curve for $\beta = 0$ corresponds to the OMIA dip shown in Fig. \ref{fig:OMIA}. We see that the only effect of finite $\beta$ is to shift the position of the OMIA dip to the left of the cavity resonance --- in the direction opposite to the shift due to increasing probe power. The shape and the depth of the dip are almost unaffected.

Next, we consider the effect of non-linear damping, adding the term $-\mu x^2 \dot{x}$ to the equations of motion for the resonator. Without the non-linear term, $\alpha=0$, this would be the van der Pol resonator \cite{Strogatz}, obeying the equation of motion
\begin{displaymath}
m_{eff} \left( \frac{d^2}{dt^2} x + (\Gamma_m - \mu x^2) \frac{d}{dt} x + \Omega^2 x \right) = F(t) \ ,
\end{displaymath}
where $F$ is an external force. For sufficiently large oscillation amplitudes the system would become unstable far from the equilibrium point ($\mu x^2 > \Gamma_m$), however, here we are interested in the situation when the non-linear term in the damping is a small correction. 

\begin{figure}
\centering
\includegraphics[width=0.4\textwidth]{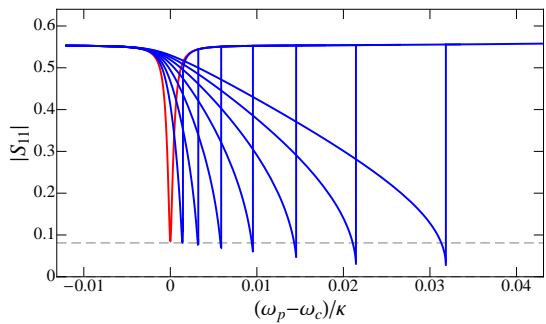}
\caption{ (Color online) OMIA response including negative non-linear damping for different probe field. The red curve shows the linear response with zero negative non-linear damping ($-\mu x^2 \dot{x}$) for the probe field $S_p=10^4  s^{-1/2}$. The dashed grey line represents the peak height of the linear response. The next blue curves include negative non-linear damping $\mu=3\times 10^{19} m^{-2} s^{-1}$. They start with the probe field $S_p=1\times 10^6 s^{-1/2}$ and end with $S_p=4\times 10^6 s^{-1/2}$ having an equally spaced step in the probe field of $0.5\times 10^6 s^{-1/2}$. The effect of the negative non-linear damping term considered here is to increase the depth of the OMIA dip in the optomechanical response with increasing driving force on the mechanical resonator.}
\label{fig:nonl_damp}
\end{figure}

Fig. \ref{fig:nonl_damp} shows the effect of the non-linear damping on the OMIA dip in the same situation as Fig. \ref{fig:OMIA}. There is an enhancement of the dip for the increasing probe power and positive $\mu$. Correspondingly, for negative $\mu$ the dip becomes deeper with increasing probe power.

\section{Conclusions}
\label{sec:conclusion}
In this Article, we systematically investigated the effect of non-linearities in a mechanical resonator on the OMIR/OMIA feature of a one-port optomechanical cavity. The main result is summarized in Fig. \ref{fig:omia_diag}. We see that depending on whether the cavity is overcoupled or undercoupled, whether it is red- or blue-sideband driven, and whether the drive power is sufficiently strong, the non-linear effects either result in OMIR with the shape repeating that of a driven Duffing oscillator (an analog of Ref. \onlinecite{Zhou}), or OMIA with a different shape, as observed in Ref. \onlinecite{exp}. The higher is the probe power, the wider is the feature and the more pronounced is the effect of non-linearity, including the width of the hysteresis range. The depth of the OMIA feature is not significally affected. Furthermore, we investigated the effect of other factors on the OMIA feature and found that the detuning of the drive shifts the OMIA dip and makes it significantly asymmetric, the quadratic term in the oscillator force shifts the position of the dip without considerably affecting its shape, whereas the non-linear dissipation enhances the dip. These conclusions will help to interpret the results of the cavity response in optomechanical cavities with a non-linear mechanical component.

Let us now discuss the limitations of our theoretical model. First, our theory is valid in the weak coupling regime, $g \ll \kappa$. Whereas this seems to be sufficient for existing experiments with non-linear resonators \cite{Zhou,exp}, multi-photon strong coupling has already been achieved in the linear mechanical limit \cite{Teufel1}. We linearize radiation-pressure interaction, and thus we can also describe multi-photon strong coupling regime using a similar framework. Generalization to the single-photon strong coupling regime is however difficult. Furthermore, we only considered the situation when the probe power is much weaker than the drive power. It would be relatively easy to consider a complementary situation when the probe is much stronger than the drive, however, we have use the method of scale separation, and it fails when the drive and the probe power are comparable. Equations still could be solved numerically in this case.

\section*{Acknowledgments}

This work was supported by the Netherlands Foundation for Fundamental Research on Matter (FOM).

\end{document}